\documentclass[letterpaper, 10 pt, conference]{llncs}

\usepackage{algorithm,algpseudocode}
\usepackage{amsmath, amsfonts}
\usepackage[utf8]{inputenc}
\usepackage{todonotes}
\usepackage{tikz}
\usepackage{cite}
\usepackage{hyperref}
\usetikzlibrary{shapes,arrows, calc, automata, fit}
\newcommand \Prob {\mathbb{P}}
\newcommand \x {\overline{x}}
\newcommand \DS {\mathit{Danger \, States}}
\newcommand \DP {\mathit{Danger \, Paths}}
\newcommand \SSt {\mathit{Safe \, States}}
\newcommand \Post {\mathit{Post}}
\newcommand \Pre {\mathit{Pre}}
\newcommand \pIC {pIC3}

\newtheorem{Lemma}{Lemma}

\begin{document}

\mainmatter  
\title{Verifying Reachability Properties in Markov Chains via Incremental Induction}
\titlerunning{pIC3}
\author{E.~Polgreen\inst{1}, M.~Brain\inst{1}, M.~Fr{\"a}nzle\inst{2}, A.~Abate\inst{1}}
\authorrunning{E.~Polgreen et al.}

\institute{ 
Department of Computer Science,  
University of Oxford \\
\and
Carl von Ossietzky University of Oldenburg    
} 

\maketitle

\begin{abstract}%4 sentences
%State the problem
There is a scalability gap between probabilistic and non-probabilistic verification. 
Probabilistic model checking tools are based either on explicit engines 
or on 
(Multi-Terminal) Binary Decision Diagrams. 
These structures are complemented in areas of non-probabilistic verification by more scalable techniques, 
such as IC3. 
We~present a symbolic probabilistic model checking algorithm based on IC3-like incremental construction of
inductive clauses to partition the state space, 
interleaved with incremental construction of a system of linear inequalities. 
This paper compares our implementation to standard quantitative verification alternatives: 
our experiments show that our algorithm is a step to more scalable symbolic verification of 
rare events in finite-state Markov chains. 
\end{abstract}

%%TEX root pic3_paper.tex
%%
%%
\section{Introduction} %1 page

Probabilistic model checking is a formal verification method used to
obtain %resulting in 
guarantees of correctness for models of real life systems that exhibit
probabilistic behaviour. 
%We study the verification of systems modelled
%as discrete-time Markov chains.
%
%Such systems are found in many
%domains~\cite{DBLP:series/natosec/KwiatkowskaP12}, both natural and engineered: 
%probabilistic behaviour may arise, for example, 
%due to failures of unreliable components, 
%due to communication across lossy media, 
%through the use of randomization in distributed protocols, 
%or due to intrinsic stochastic behaviour in biological models. 
%
Current verification engines based on sparse matrices, BDDs, and MTBDDs are fast
but scalability is dependent on the structure
of the model~\cite{DBLP:journals/sttt/KwiatkowskaNP04}. 
%and there is a need for a more universally increased scalability across
%all model structures. 
%
%\todo[inline]{this example still needs changing}
%\input{example}
%
%
In non-probabilistic verification recent
SAT/SMT-based engines, such as Craig
interpolation~\cite{DBLP:journals/tcs/McMillan05} or
IC3~\cite{DBLP:conf/vmcai/Bradley11}, complement BDD-based model checking; whilst
often slower, these engines are able to tackle some models where BDDs require too much memory. 
This paper presents an algorithm for the analysis of probabilistic safety/reachability
properties based on incrementally constructing inductive invariants, a principle introduced in
IC3~\cite{DBLP:conf/vmcai/Bradley11}, and incrementally constructing a system of linear inequalities.

IC3 is a SAT-based, yet complete, algorithm that computes an
over-approxima\-tion of the set of reachable states 
in the form of an inductive invariant that implies a safety property.
%
%IC3 develops bounded model checking further, and is purely SAT-based, yet
%complete. 
%A key benefit of IC3 is that it only uses one copy of the transition relation,
%and is thus memory efficient, in comparison to Bounded Model Checking, which
%requires $n$ copies of the transition relation to find counter-examples of
%length $n$. BDDs only require one copy of the transition relation, but with
%large models the BDDs become too large for even just one copy.
%
One of the challenges in applying inductive invariants to probabilistic properties is that there
is no direct equivalent
of this invariant for probabilistic safety properties, since these are 
properties that depend on multiple execution paths rather than just reachable
states. In other words, a counter-example for a non-probabilistic safety property is 
any single path from the initial state to the unsafe states;
a counter-example for a probabilistic safety property is a group of paths from the initial to
the unsafe states with an accrued probability greater than a specified threshold.
Existing work uses SMT techniques to produce counter-examples for probabilistic 
systems as sets of paths~\cite{DBLP:conf/forte/BraitlingWBJA11}, or as 
minimal critical subsystems~\cite{DBLP:conf/tacas/WimmerJABK12}, but these techniques are based on Bounded Model Checking(BMC). There are several advantages to using IC3 over BMC~\cite{DBLP:conf/cav/CimattiG12}: first it is only necessary to unroll the transition relation one step; second, the incremental construction of invariants allows it to make much more effective use of incremental SAT solving than BMC; and finally the reasoning in IC3 is localised and driven by the property being checked. 
We exploit a by-product of IC3's invariant construction to incrementally build a system of 
linear inequalities based on the set of $\DS$,
defined as the set of states that may eventually lead to a violation of the
safety property (i.e., reach the unsafe state). This is an over-approximation of the minimal
critical subsystem.
The system of linear inequalities represents upper and lower bounds on the probability of
violating the safety property for states in this set. At each iteration of the algorithm, we check 
whether these are sufficient to decide the validity of the property.  

The technical contributions in this paper are: \vspace{-0.1in}
%\mbcomment{Listing contributions is good,linking them to Sections is even
%better}. 
\begin{itemize} \item In Section
\ref{sec:alg} we present an algorithm for verifying reachability properties in Markov chains, which uses incrementally constructed inductive
invariants to partition the state space and incrementally constructs a
system of linear inequalities bounding the reachability probability. 
We use SMT techniques to establish whether the reachability probability falls
within the
acceptable range or not. We call our algorithm pIC3, as it is a probabilistic model checking technique inspired by IC3.
\item We prove the soundness and termination of the
algorithm in Appendix \ref{sec:proof}. 
\item In Section \ref{sec:implement} we describe the implementation of pIC3, and
explore its scalability. We find that the scalability of pIC3 is dependent on
the number of $\DS$ in the model, and that pIC3 can verify rare-events on models
where PRISM fails.
%compare its performance to standard probabilistic model checking algorithms, 
%as implemented in the PRISM model checker \cite{DBLP:conf/cav/KwiatkowskaNP11}. 
\end{itemize} 

%%TEX root pic3_paper.tex
%%
\section{Background}

\subsection{Discrete Time Markov Chains}

The probabilistic transition systems we will consider are given as Discrete
Time Markov Chains (DTMC)~\cite{DBLP:books/daglib/0020348}. We consider symbolic DTMCs 
represented as follows:
$D = (\x, I(\x), T(\x, \x'), P(\x, \x'))$, where
\begin{itemize}
\item $\x = \{x_0, ..., x_n\}$ is a finite set of Boolean state variables.  Each
complete assignment of Boolean variables corresponds to a single state, $s$. We
denote a single state $s$ and its valuation $\x$ by $s(\x)$: 
e.g., for $\x = \{x_0, x_1\}$ and $s = \neg x_0 \wedge x_1$, we denote the state
by $s(01)$.
\item we write $\x'$ to denote the set of next state Boolean variables. 
Applying prime to a formula $F$ is the same as applying prime to all of its
variables.
\item $I(\x)$ is a predicate over the state variables, representing the initial
states. We assume only one initial state, but will outline an extension to
multiple initial states in Section \ref{sec:extensions}.
\item $T(\x,\x')$ is a transition relation over a set of state variables $\x$,
that holds if there is a transition from $\x$ to $\x'$.
\item $P(\x, \x')$ maps a transition to a probability in $[0,1]$, e.g., $P(x_0
\wedge \neg x_1,\neg x_0 \wedge x_1)$ gives the probability of a transition
occurring from the state given by $x_0 \wedge \neg x_1$ to the state given by
$\neg x_0 \wedge x_1$
\footnote{In this work we use $P$ to denote transition probabilities, $Pr$ to
denote probabilities associated to 
%events or 
paths, and $\mathbb P_{(\cdot)}$ for the probabilistic operator in PCTL
formulae.}.
 If $\x$ or $\x'$ are not full assignments, $P$ maps several transitions to the
same probability, allowing symbolic storage of the transition probabilities. If
a state $s(\x)$ has any incoming or outgoing transition, i.e., there exists a
satisfying assignment to $T \wedge s$ or $T \wedge s'$, the outgoing transition
probabilities of the state must sum to 1.
\end{itemize}
It is not strictly necessary to define both $T(\x,\x')$ and $P(\x, \x')$, as
$T(\x,\x')$ evaluates to true if $P(\x, \x')>0$, 
and to false otherwise. 
Retaining both is practical for implementation purposes. 
We use $S$ to denote the finite state-space, which is the set of all possible complete
valuations of $\x$.

\smallskip

%\paragraph{}
We will describe our method using a small example transition system, illustrated in
Figure~\ref{fig:step1}.
However, our algorithm is designed to handle transition systems that are too
large
to be practically represented explicitly. 
The example has two state variables, $x_1$ and $x_2$, and we represent each
possible combination as an explicit state encoded with $s(x_1x_2)$. The 
property is $\phi = \neg x_1 \vee \neg x_2$ and the initial states are given by
$I = \neg x_1 \wedge \neg x_2$. 

\subsection{Probabilistic safety properties}
\label{sec:property}

We consider unbounded horizon safety properties with a probabilistic bound,
e.g., the probability of reaching a failure state
must be less than $y$. Formally, we write $\Prob_{< y}[\Diamond \neg\phi \mid I]$,
where $y \in [0,1]$, $\neg\phi$ is a predicate over the state variables
that represents the set of failure states, and $I$ is the set of initial states. By abuse
of notation we use $\Prob_{< y}[\Diamond \neg\phi]$ to denote this property starting from
the initial states.

%A probabilistic
%%safety property is evaluated by considering all path fragments that originate in an
%initial state and finish in a failure state,
%and finding the sum of the probabilities for each of those paths occurring. We
%call these path fragments $\DP$.
%\paragraph{} Using the example in
%Fig.~\ref{fig:step1}, we specify $\phi = \neg x_1 \vee \neg x_2$ and ask whether
%$\Prob_{<0.5}[\Diamond \neg \phi]$. The $\DP$ are those paths that start
%in $s(00)$ and finish at $s(11)$, 
%the only state satisfying $\neg \phi$. The paths that match this specification
%take the form, 
%$s(00)(s(10))^+ s(11)$, and have associated likelihoods $0.5 \cdot (0.5)^+ \cdot
%0.2$, 
%which sum to a probability equal to $0.1$, 
%and thus the property holds. 

\subsubsection{Property as path probabilities}

A probabilistic
safety property is evaluated by considering all finite paths that originate in an
initial state and finish in a failure state,
and finding the sum of the probabilities for each of those paths occurring. We
call these path fragments $\DP$.
The property $\Prob_{<y}[\Diamond \neg \phi | I]$, can be written in terms of
the sum of probabilities of each of the $\DP$ occurring.

%We denote a finite path fragment in our DTMC $D$ by $\omega \in \mathit{Paths}$, and
%the set of paths possible starting from a state $s$ by $\mathit{Paths}(s)$. 
%
A $\mathit{Danger \,Path}$ is defined as a finite path that starts in the initial
state, and eventually leads to a failure state, i.e., it satisfies
$\Diamond \neg\phi$,  
\begin{equation*}
\omega_{fin} \models \Diamond \neg \phi \iff \exists i \geq 0. \, (\omega(i)\models
\neg \phi).
\end{equation*}

\noindent The probability
of a finite path of length $n$, $\omega_{fin}^n$ occurring,
$Pr(\omega_{fin}^n)$, is defined as:
\begin{equation*}
Pr(\omega_{fin}^n) = P(\omega(0),\omega(1))\cdot ... \cdot P(\omega(n-1),
\omega(n)).
\end{equation*}
Now our property can be written in terms of the sum of the path probabilities
for all the $\DP$: 
%\begin{equation*}
$Pr[\Diamond \neg \phi \mid I] = \sum_{\forall \omega_{fin} \models \Diamond \neg
\phi} Pr (\omega_{fin})$. 
\noindent Note that, for the $\DP$,
$\omega(0), ..., \omega(i-1) \in (\Post^*(I) \cap \Pre^*(\neg \phi))$. 
%We will use this result.
%
\subsection{Bounds on path probabilities}

We use the notation $U_p(s)$ to denote an upper bound for the $Pr[\Diamond \neg
\phi \mid s]$, i.e., an upper bound on the probability of reaching a $\neg
\phi$ state starting from $s$.
We use $L_p(s)$ to denote a lower bound for the $Pr[\Diamond \neg
\phi \mid s]$, i.e, a lower bound on the probability of reaching a $\neg \phi$
state starting from $s$.

\subsection{IC3} %\mbcomment{move this paragraph to the background section} 
IC3~\cite{DBLP:conf/vmcai/Bradley11} proves or refutes safety properties 
%i.e., properties of the form $\Diamond \neg \phi$, 
by incrementally constructing an inductive invariant that implies $\phi$. 
It does this by building a
series of frames, i.e., sets of states that over-approximate the states
reachable within $0,...,k$ steps. 
%
%Starting with $F_0 = I$ and $F_1 = \phi$, 
IC3 removes states in the
frames that can reach $\neg\phi$, known as \textbf{counterexamples to
induction} (CTIs), and proves or disproves their reachability in a backward-search
fashion. 
%These states are called ``counter-examples to induction'' and are states that
%can reach a ``bad state'' in one step, i.e., they have a transition to a state
%in $\neg \phi$ or a transition to a state that is known to be able to reach
%$\neg \phi$.
It continues this process until it either finds a counterexample path from the initial states
to $\neg \phi$ or
until it obtains a frame closed under the transition relation, in which all
states satisfy the property. 
By this technique IC3 avoids the explicit unrolling of
the transition relation that is required by bounded model checking and explicit
reachability analysis. 

\subsubsection{Conjunctive Normal Form and Frames}

A formula is in conjunctive normal form (CNF) if it is a conjunction of clauses,
where each clause, $c$, is a disjunction of literals.
IC3 tightly integrates
with a propositional SAT solver, which requires Boolean formula in CNF.

A frame $F$ is a formula in conjunctive normal form that represents a set of
states in the state space, i.e., $F_i\subseteq S$, and is a conjunction of
multiple clauses.

\subsubsection{Induction}
A frame $F$ is an \emph{inductive invariant} for a transition system $S$ if
$I\implies F$ and $F \wedge T \implies F'$, i.e, if all initial states are
covered by $F$ and $F$ is closed under the transition relation. A clause $c$ is
inductive relative to a frame $F$ if both $I \implies c$ and $F \wedge T \wedge
c \implies c'$.

\subsubsection{Post and Pre}
We use the notation $\Post^n(s)$ for the set of states that are
reachable from $s$ within $n$ steps using transitions in $T$, and $\Post^*(s)$
for the set of states that are
reachable from $s$ using transitions in $T$. Similarly, we use the
notation $\Pre^n(s)$ for the set of states that can reach $s$ within $n$ steps
using transitions in $T$, and $\Pre^*(s)$ for the set of states that can reach
$s$ using transitions in $T$~\cite{DBLP:books/daglib/0020348}.

%\missingfigure{small example adapted from Somenzi's IC3 talk, Fig 1}
\subsubsection{Paths}
A path through a DTMC is a sequence of states $s_0s_1s_2s_3 ...$, starting from
an initial state, such that $\forall i. \, P(s_i, s_{i+1})>0 $, and represents
an execution of the system that the DTMC models. We denote a path by $\omega$, and
a
finite path by $\omega_{\mathit{fin}}$, 
and $\mathit{Paths}(s)$ $(\mathit{Paths}_{\mathit{fin}}(s))$ is the set of all
infinite (finite) paths in the DTMC $D$ starting from state $s$. We use $\omega(i)$ to denote
the $i^{th}$ state in a path.

\section{Our Algorithm (pIC3)}
\label{sec:alg}
We have noted in Section~\ref{sec:property} that a probabilistic reachability
property can be expressed in terms of the sum of the path probabilities of all
$\DP$, 
and that all $\DP$ are contained within $\Post^*(I) \cap
\Pre^*(\neg\phi)$.
PIC3 therefore incrementally (by introducing and refining successive frames
$F_0, ..., F_n$) identifies three sets of states within the state space, shown
in Figure~\ref{fig:statespace}:
\begin{itemize} 
\item $\neg\phi$; 
\item $\SSt$, which are states from which $\neg\phi$ cannot be reached;  
\item $\DS$, which are states in $\phi$ that may reach states in $\neg \phi$.
This set includes, at termination, $\Post^*(I) \cap \Pre^*(\neg\phi)$, i.e., all
the $\DP$. After the $nth$ frame has been added, $\DS$ includes $\Post^n(I)\cap
\Pre^n(\neg\phi)$, i.e., all the $\DP$ of length $n$ or less.% plus the first
\end{itemize}

%\begin{figure}
%\begin{center}
%\begin{tikzpicture}
%
%\draw  (0,0) rectangle node {$\Post^*(I) \cap \neg\Pre^*(\neg\phi) \subseteq
%\SSt$} (7,2);
%\draw (0,2) [fill = lightgray]rectangle node [align=center] {$\Post^*(I) \cap
%\Pre^*(\neg\phi) \subseteq \DS$}(7.5,4);
%\draw(7,0) rectangle node {$\neg \phi$}(9,4);
%\draw[red,thick,dashed](-0.1,-0.1) rectangle (7.1,4.1);
%\node [draw, red] at (3.5, -0.3) {$\phi$};
%
%\end{tikzpicture}
%\end{center}
%\caption{Partitions of the state space of the model.\label{fig:statespace}}
%\end{figure}
%\todo[inline]{overlapping rectangles between DS and $\neg\phi$}
%
%
We begin with a state space specified by state variables, and the states that
violate the property are identified by the boolean formula $\neg \phi$. We
symbolically partition out the $\SSt$ by incrementally constructing inductive invariants, 
and we identify $\DS$ as the states in $\phi$, but not in $\SSt$. 
After each frame is added pIC3 checks whether the system of linear inequalities we
have so far constructed from states found so far in $\DS$ is sufficient to prove or refute
the property.  \\

%\paragraph{}
To improve efficiency, we store only a partial explicit list of $\DS$, namely only the
CTIs explicitly identified by the SAT solver calls, from which we construct the system
of linear inequalities,
and we make SMT calls to check whether the system of linear inequalities contains enough information to
prove or refute the property. 
We increase the number of states and transitions in this explicit list and system of linear inequalities until we
can prove or refute the property. Note that, although the SAT solver returns
explicit instances of $\DS$ and we store this explicit list, we in fact make a
generalisation about each explicit state and remove both the state and a group
of similar states from $\SSt$. 
Thus $\DS$ is composed of two parts: the explicit list, plus the symbolically
handled set $\phi \setminus \SSt$.
%\vskip 1em
%$S \setminus (\SSt \cup \neg \phi)$. 
%This explicit list is equivalent to the list of states that IC3 explicitly
%identifies as counterexamples to induction, referred to as proof
%obligations~\cite{DBLP:conf/vmcai/Bradley11}. 
%

\noindent The two key processes of pIC3, that are interleaved, are:
\begin{itemize}
\item we partition the state space symbolically, by incrementally constructing inductive invariants; 
\item we incrementally collect information about the probability of the $\DP$,
and the algorithm terminates when we have collected the minimum information that
is sufficient to prove or refute the given property $P_{>y}
[\Diamond \neg\phi]$. 
\end{itemize}

In this section we will first give an overview of the partitioning out of the
$\SSt$, and relate this back to the original IC3 algorithm. Then we will explain
the $\DS$ collection and system of linear inequalities, and how it interleaves into the state-space 
partioning. Finally, we will explain
how, after discovery of new $\DS$, we make decisions about the decidability of
the property based on system of linear inequalities constructed so far. A~block diagram of the
algorithm is given in Figure~\ref{fig:block_diag}. The full pseudo-code is found
in Appendix~\ref{sec:proof}. 

\subsection{Partitioning out $\SSt$}

\paragraph{IC3} generates an invariant that implies a property $\phi$ by
constructing sets of states, called \textbf{frames}, $F_0, ..., F_k \subseteq
S$, that are over-approximations of states reachable within $0, ..., k$ steps.
It terminates when it finds a path that leads from the initial state to
$\neg\phi$, in which case the property does not hold, 
or when it reaches a frame that is closed under the transition relation and does
not contain a state in $\neg\phi$. In this case the property holds, and the
frame is an invariant that implies the property. 
IC3 builds the frames by starting with the largest possible over-approximation
for a frame, and then uses a SAT solver to find counterexamples to this as an
invariant that implies the property, i.e., states in the frame that have a
transition that leads to $\neg \phi$ in one step. For each frame, it removes any
of these states, returned by the SAT solver, by adding clauses that imply the
negation of these states. 
Each time a new frame is added, any clauses that are relatively inductive, i.e.,
hold in both a frame and in the next frame, are propagated forward to the new
frame.

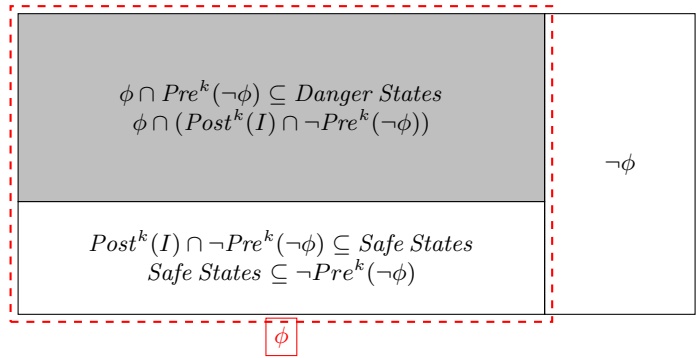
\begin{figure}[h]
\begin{center}
\begin{tikzpicture}

\draw (0,0) rectangle node [align=center] {$\Post^k(I) \cap \neg\Pre^k(\neg\phi)
\subseteq \SSt
 $\\ $\SSt \subseteq \neg Pre^k(\neg\phi)$} (7,1.5);
\draw (0,1.5)[fill=lightgray] rectangle node [align=center] {$\phi \cap
\Pre^k(\neg\phi) \subseteq \DS$\\
$\phi \cap (\Post^k(I) \cap \neg\Pre^k(\neg\phi) ) $}(7,4);
\draw(7,0) rectangle node {$\neg \phi$}(9,4);
\draw[red,thick,dashed](-0.1,-0.1) rectangle (7.1,4.1);
\node [draw, red] at (3.5, -0.3) {$\phi$};

\end{tikzpicture}
\caption{Partions of the state space after $k$ steps. \label{fig:statespace}}
\end{center}
\end{figure}

\paragraph{pIC3} defines $\SSt$ as the states that can never reach $\neg \phi$,
and attempts to build an invariant that implies $\SSt$. This invariant
represents an over-approximation of $Post^*(I) \cap \neg \Pre^*(\neg \phi)$. We
construct the invariant by building a series of frames $F_0, ..., F_k \subseteq
S$, which represent an over-approximation of the set of states reachable within
$0, ..., k$ steps that can not reach $\neg\phi$ within $k$ steps, formally
$(\Post^k(I) \setminus \Pre^k(\neg \phi)) \subseteq F_k$. At the end of the
$kth$ iteration of the main loop, the state space will be partitioned as shown
in Figure~\ref{fig:statespace}.

%\vskip -0.5cm

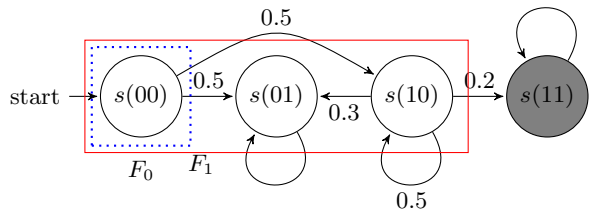
\begin{figure}[h]
\begin{center}
\begin{tikzpicture}[>=stealth',shorten >=1pt,auto,node distance=1.8cm, scale =
1, transform shape]
\node[initial,state] (A) {$s(00)$};
\node[state](B) [right of= A]{$s(01)$};
\node[state](C) [right of = B]{$s(10)$};
\node[state, fill=gray](D) [right of = C]{$s(11)$};
\path[->] 
	(A) edge  node {$0.5$}(B)
	(A) edge [bend left, min distance=15mm] node {$0.5$}(C)
	(C) edge node {$0.3$}(B)
	(C) edge node {$0.2$}(D)
	(B) edge [loop below, min distance = 10mm, in=240,out=300, looseness=5] node
{}(B)
	(C) edge [loop below, min distance = 10mm, in=240,out=300, looseness=5] node
{$0.5$}(C)
	(D) edge [loop above, min distance = 10mm, in=120,out=60, looseness=5] node
{}(D);
	\node [draw=red, fit= (A) (B) (C), inner sep=0.2cm, label={[xshift = -1cm,
yshift=-1.9cm]$F_1$}] {};
	\node [draw=blue, thick,dotted, fit= (A), label={[yshift=-1.9cm]$F_0$}] {};
\end{tikzpicture}
\caption{Initialisation of $F_0 = \neg x_1 \wedge \neg x_2$ and $F_1 = \neg x_1
\vee \neg x_2$. \label{fig:step1}}
\end{center}
\end{figure}

We will illustrate our method on the simple explicit 4-state system, shown in
Figure~\ref{fig:step1}. 
We initialise the frames with $F_0 = I$, as this is exactly the set of states
reachable within $0$ steps, and $F_1 = \phi$, as this is the biggest
approximation possible of states reachable within $1$ step that do not violate
$\phi$. Frames are constructed and modified via two processes:
propagation of inductive clauses, and frame refinement by removal of violating
states and generalisations of the violating states. We will first talk about
frame refinement. 
%\vskip -0.5cm

\subsubsection{Frame refinement}
Consider that we have added a frame $F_k$, which is the best over-approximation
we have of the $\SSt$ reachable within $k$ steps. In this stage of the
algorithm, we remove any state $s \in F_k$ that can reach $\neg\phi$ within one
step via the transition relation $T$. For each state $s \in F_k$ that can reach
$\neg \phi$ in one step, we remove all states $t \in F_{k-1}$ that can reach $s$
within one step via $T$, and for each state $t \in F_{k-1}$, we remove all
states $r \in F_{k-2}$ that can reach $t$ within one step via $T$, and so on.
The states we remove are CTIs. By this method, we
can remove all states in $Pre^k(\neg \phi)$.

\begin{figure}[h]
\begin{center}
\begin{tikzpicture}[>=stealth',shorten >=1pt,auto,node distance=1.8cm, scale =
1, transform shape]
\node[initial,state] (A) {$s(00)$};
\node[state](B) [right of= A]{$s(01)$};
\node[state](C) [right of = B]{$s(10)$};
\node[state, fill=gray](D) [right of = C]{$s(11)$};
\path[->] 
	(A) edge  node {$0.5$}(B)
	(A) edge [bend left, min distance=15mm] node {$0.5$}(C)
	(C) edge node {$0.3$}(B)
	(C) edge node {$0.2$}(D)
	(B) edge [loop below, min distance = 10mm, in=240,out=300, looseness=5] node
{}(B)
	(C) edge [loop below, min distance = 10mm, in=240,out=300, looseness=5] node
{$0.5$}(C)
	(D) edge [loop above, min distance = 10mm, in=120,out=60, looseness=5] node
{}(D);
	\node [draw=red, fit= (A) (B), inner sep=0.2cm,
label={[yshift=-1.9cm]$F_1$}] {};
	\node [draw=blue, thick,dotted, fit= (A), label={[yshift=-1.9cm]$F_0$}] {};
\end{tikzpicture}
\caption{Elimination of state $s(10)$ from $F_1$.\label{fig:step2}}
\end{center}
\end{figure}
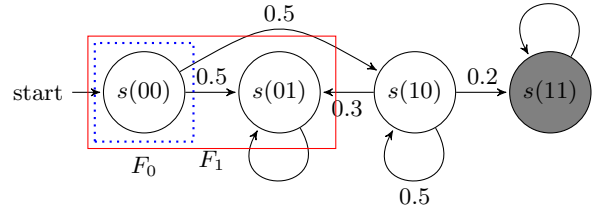

In Figure~\ref{fig:step2}, the SAT solver has returned the one-step transition
from $s(10)$ to $s(11)$ and so we remove state $s(10)$ from frame $F_1$. 
%\vskip -0.5cm
%\vskip -0.5cm

We remove the states by adding boolean clauses to the frames that imply the
negation of the state, i.e., to remove state $s$, we add a clause $c \implies
\neg s$. We~could use $c = \neg s$, but,
in fact, we hope to be able to find a sub-clause of $\neg s$, which is inductive
relative to our frames and will remove a set of $s$-like states. We elaborate this
step next.

\subsubsection{Clause generalisation}
\begin{figure}
\begin{center}
\begin{tikzpicture}[>=stealth',shorten >=1pt,auto,node distance=1.8cm, scale =
1, transform shape]
\node[state,minimum size=1.2cm, white]  (A) {};
\node [state,minimum size=1.2cm]  (B) [right of =A] {$s(110)$};
\node [state,minimum size=1.2cm] (D) [right of =B] {$s(001)$};
\node [state,minimum size=1.2cm] (C) [right of =D] {$s(101)$};
\node [state,minimum size=1.2cm]  (F) [right of =C]{$s(100)$};
\node [state,minimum size=1.2cm, fill=gray]  (E) [right of=F] {$s(111)$};

\path[->]
	(B) edge (D)
	(D) edge (C)
	(D) edge [loop below, min distance = 10mm, in=240,out=300, looseness=5](D)
	(C) edge [bend left, min distance=15mm] (E)
	(E) edge [loop above, min distance = 10mm, in=120,out=60, looseness=5] (E)
	(F) edge (E);
	\node [draw=red, fit= (A) (B) (C) (D) (F), inner sep=0.4cm, label={[xshift =
2.5cm, yshift=-2.1cm]$F_k$}] {};
	\node [draw=blue, thick,dotted, fit= (A) (B) (D),inner sep=0.2cm,
label={[yshift=-1.7cm, xshift=-0.6cm]$F_{k-1}$}] {};
	\node[draw=black, fill=lightgray, fit=(A), label={[yshift=-1.5cm]$F_{k-2}$}]{};
 
\end{tikzpicture}
\caption{Example of generalisation procedure on a fragment of a DTMC. \label{fig:gen}}
\end{center}
\end{figure}
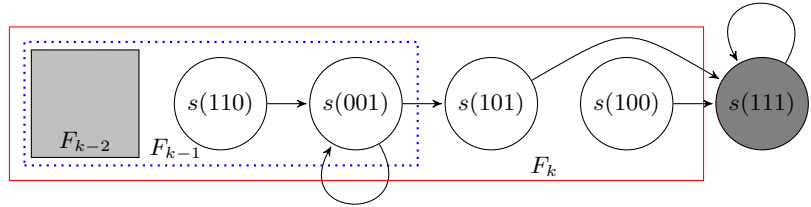

Consider Figure~\ref{fig:gen}, which shows a fragment of a DTMC (the states inside frame $F_{k-2}$ 
and the states that have already been removed from frames are not shown).
Suppose the SAT solver returns the state
$s(100)$. The negation of this state, $\neg x_1 \vee x_2 \vee x_3$ is inductive
relative to $F_0, ..., F_k$, because the state has no predecessors. We wish to
generalize from this information, and so we look to see if there are variables
in the clause that can be inverted while the clause remains inductive, and thus
can be assigned ``don't care''. Consider $\neg x_1 \vee x_2 \vee \neg x_3$. This
is not inductive relative to $F_{k-1}$ because state $s(101)$ has a predecessor
in frame $F_{k-1}$, so we cannot drop variable $x_3$. Now consider $\neg x_1
\vee \neg x_2 \vee x_3$. This is inductive relative to $F_0, ..., F_k$ because
the state $s(110)$ has no predecessors. We can hence use the clause $c = \neg
x_1 \vee x_3$.
%\vskip -0.5cm
%\vskip -0.5cm

The effectiveness of the generalisation depends on the connectivity of the state
graph and its encoding, and is key to IC3 being able to handle large state
spaces symbolically. The original IC3 algorithm implements a generalisation
procedure that finds a \emph{minimal} inductive
clause~\cite{DBLP:conf/fmcad/BradleyM07}, and further work improves this by
extracting information from the counterexamples to
generalization~\cite{DBLP:conf/fmcad/HassanBS13}.  
We implement a simple ``greedy'' generalization scheme for \pIC, where we
attempt to drop each variable of the clause in sequence, and drop the first one
that gives an inductive sub-clause. This approach is sub-optimal,
but is nevertheless better than taking $c=\neg s$ and removing only one state at
a time.

\subsubsection{Clause propagation}
When we add the first frame $F_1$, we take the largest possible
over-approximation because we have no additional information. For all subsequent
frames, we have additional information in the form of states we have removed
from previous frames. When we add a frame $F_{k+1}$, we start with the largest
over-approximation of $F_{k+1} = \phi$, and then we look one by one at the
clauses in the frames $F_1,..., F_k$ and see whether they can be propagated
forward to the next frames. We propagate relatively inductive clauses, i.e., a
clause $c$ in $F_i$ that holds in $F_{i+1}$ will be added to $F_{i+1}$. 
%We establish whether a clause is relatively inductive by asking a SAT solver if
%there exists a transition from $F_i \wedge c$ to a state that does not satisfy
%$c$ via the transition relation $T$. 
%
Note that if we have added clauses in the frame refinement stage that eliminate
states that have predecessors and so potentially are part of $Post^*(I)$, the
clauses will fail to propagate to the new frame, until $k$ is big enough that we
have found all the predecessors.%, or we have found that the states are part of

\subsection{Building $\DS$}
\label{sec:buildingDS}
In the previous section we have described how by making SAT calls we find states
that are in $Pre^*(\neg\phi)$. 
When the SAT solver returns a CTI in the form of an
explicit state and transition to a successor state, we remove the state from $\SSt$, as
described above, and we add both states to an explicit list of $\DS$. As we find each
state we incrementally construct an SMT problem made up of upper and lower bounds
for $Pr[\Diamond \neg \phi \mid s]$ $\forall s \in
\DS$ with the information we have about $\DS$. These upper and lower bounds 
are a system of linear inequalities derived from the partial knowledge of the system of 
linear equations that is used to calculate reachability probabilities in 
existing probabilistic model
checking algorithms~\cite{DBLP:books/daglib/0020348}.

When we find that $I$ is in $Pre^*(\neg\phi)$, i.e., we have found at least one
complete path in $\DP$, we check whether we have enough information to prove or
refute the property yet, i.e., whether all solutions to $Pr[\Diamond \neg \phi \mid I]$ are
guaranteed to be one side of the probability threshold.

\subsubsection{Probability bounds}
\label{sec:bounds}
In Section~\ref{sec:property} we showed that $Pr[\Diamond \neg \phi \mid I]$ can
be expressed as path probabilities. We can extend this result to write upper and
lower bounds, $U_p(s)$ and $L_p(s)$, for $Pr[\Diamond \neg \phi \mid s]$ in
terms of path probabilities. The lower bound can be written as follows:
\begin{equation*}
L_p(s) = \sum_{t \in \DS \setminus \neg \phi} P(s,t). L_p(t) + \sum_{u \in \neg
\phi}P(s,u). 
\end{equation*}
In other words, either $\neg \phi$ is reached within one step, i.e., by
a transition $s$ to $u \in \neg \phi$, or by a transition to a state $t\in
S\setminus \neg \phi$, from which a path to $\neg\phi$ is
taken. This corresponds to all path fragments,
where all states (except the last one) are not part of $\neg
\phi$~\cite{DBLP:books/daglib/0020348}. There is a unique solution to $L_p(I)$,
because all states in $\DS$ are backward reachable from $\neg\phi$, and all
states in $\neg\phi$ have $L_p(s)= U_p(s) = 1$. 

The sum of the probabilities on transitions leaving any state must equal one, so
we can obtain an upper bound for all $s \in$ $\DS$. 
%The upper bound is another assertion added to the SMT solver, and appended to
%when a new transition from $s$ is found. 
There is a unique solution to $U_p(I)$, because there is a unique solution to
$L_p(s)$.
\begin{equation*}
 U_p(s) = L_p(s) + 1 - \sum_{\forall s' \in S}T(s,s'). 
 \end{equation*}

We do not explicitly compute the values for $L_p(s)$ and $U_p(s)$, but 
instead incrementally construct the SMT problem described above, adding more
constraints as we find each transition. This allows us to 
make use of the incremental solving capabilities of Z3. Representation of
the upper and lower bounds as a system of linear inequalities allows compact
representation of paths that contain loops, or multiple paths containing the same
states and transitions.

%We do not explicitly compute the values for $L_p(s)$ and $U_p(s)$ each time we
%discover a new transition.
%Instead, we map each member of our explicit list of $\DS$ to a vector of
%successor states. Transition probabilities are extracted from the symbolically
%stored transition matrix. We append to this list when new transitions are found,
%and 
%assertions for each upper and lower bound are constructed from this list and
%sent to the SMT solver for further use (see below).  
%The list of successor states for each $\DS$ only includes transitions we have
%explicitly found with the SAT calls and so is incomplete until final
%termination. 

Note that the SMT problem includes only transitions we have explicitly found with SAT
calls, and so is incomplete unless we reach final termination. 

In Figure \ref{fig:step2}, the SAT solver has returned the transition from
$s(10)$ to $s(11)$. 
$P(s, s')$ gives the transition probability $p$ which in our example is equal to
$0.2$. 
We now know that the lower bound for $Pr[\Diamond \neg \phi \mid s]$ is the
transition probability $p$ multiplied by the lower bound for $Pr[\Diamond \neg
\phi \mid s']$, i.e., $0.2 \cdot L_p(s(11))$. 
We also know the upper bound for $Pr[\Diamond \neg \phi \mid s]$ is $p \cdot
Pr[\Diamond \neg \phi \mid s'] + (1-p)$, because the worst-case scenario is that
all remaining, not yet found, transitions from $s$ lead directly to $\neg\phi$.
In our example $U_p(s(01)) = 0.8 + 0.2 \cdot U_p(s(11))$. 

We make two calls to the SMT solver: 
the first one checks whether the reachability probability is guaranteed below
the threshold $y$, by adding the assertion $U_p(I)<y$; 
the second call checks whether the reachability probability is guaranteed above
the threshold $y$, by adding the assertion $L_p(I) \geq y$. 

If $U_p(I)<y$ is not satisfiable then $Pr[\Diamond \neg\phi \mid I]\geq y$, 
whereas if $L_p(I)\geq y$ is not satisfiable $Pr[\Diamond \neg\phi \mid I]<y$
and the property holds. We call both of these cases ``early termination''.

However, if we have found the complete set of $\DS$ and all transitions between
them, then the lower bound will equal the upper bound for every state: 
in this case the reachability property is the unique solution to the system of
linear equations. We call this ``final termination''. 

%Note that this representation of reachability probabilities is used in existing
%probabilistic model checking algorithms~\cite{DBLP:series/natosec/Katoen13,
%DBLP:books/daglib/0020348}.

\subsubsection{Detecting loops}
We use incremental SAT solving to find any not-yet-found transitions in $T$
between explicitly known states in $\DS$, thus finding any loops in the $\DP$ we have found so far. 
We may choose the frequency with which to run this 
search, e.g., every iteration, or only every $N^{th}$ iteration, or we may 
omit this step completely and find the loops after
we have found an invariant that implies $\phi \setminus
\Pre^*(\neg\phi)$. In our experimental evaluation, we do the latter, as we find it gives marginally
quicker run times with the model we are using. 

\subsubsection{Adding information from $\DS$ to the system of linear inequalities}

IC3 terminates when it has produced an invariant that implies $\phi$, or it has
found a counterexample path from $I$ to $\neg\phi$. 
pIC3 terminates when it has found enough $\DS$ to prove or refute the property. 
It is possible that pIC3 may find an invariant that implies $\phi \setminus
\Pre^*(\neg\phi)$ before this happens. At this point, we know we have found the
full symbolic set of $\DS$, but we may not have added enough information about
the transitions within $\DS$ to our system of linear inequalities. 

In this case, pIC3 uses incremental SAT solving to extract information from the symbolically represented
$\DS$, in the form of transitions in $T$ that we have not yet discovered, and adds the information
 to the system of linear inequalities.
We do this until we have enough information to terminate. 

If we have added every transition within $\DS$ to the system of linear inequalities, 
we have a full set of linear equations to
solve where $U_p(s)=L_p(s)$ for every state. This will give us an exact answer
for $Pr[\Diamond \neg\phi \mid I]$. It is, however, unlikely that we will reach
this full set of linear equations without being able to prove or refute the
property at an earlier stage.

\begin{center}
\begin{figure}
\tikzset{
    state/.style={
           rectangle,
           rounded corners,
           draw=black, very thick,
           minimum height=2em,
           inner sep=2pt,
           text centered,
           },
}

\begin{center}
\begin{tikzpicture}[->,>=stealth']
 % Position of QUERY 
 % Use previously defined 'state' as layout (see above)
 % use tabular for content to get columns/rows
 % parbox to limit width of the listing
 \node[state, text width=3cm] (REFINE) 
 {\textbf{Refine frames:}\\
 remove $Pre^k(\neg\phi)$ from $F_k$ and add to DangerStates.

 };

  \node[state, initial, above of = REFINE, node distance=2cm] (INIT) 
 {Initialise $F_0 = I$, $F_1 = \phi$};
  
 % State: ACK with different content
 \node[state,    	% layout (defined above)
  text width=3cm, 	% max text width
  right of=REFINE, 	% Position is to the right of QUERY
 node distance=3.5cm, 	% distance to QUERY
  anchor=center] (CHECK) 	% posistion relative to the center of the 'box'
 {%
 \textbf{Check probability bounds:}
 is $P[\Diamond \neg\phi \mid I]<y$? 
 };

 \node [state, right of =CHECK, node distance =3cm, yshift=0.5cm](YES){PASS};
 \node [state, right of =CHECK, node distance =3cm, yshift=-0.5cm](NO){FAIL};

 % STATE QUERYREP
 \node[state,
  below of=REFINE,
  node distance=4cm,
  anchor=center,
  text width=3.5cm] (PROP) 
 {\textbf{Propagate clauses:} Add new frame and propagate clauses forward\\ $F_k  = F_{k+1}$?
 };

 \node[state,text width=3cm,
  below of=CHECK,
  node distance=2cm,
  anchor=center] (LOOPS) 
 {Detect loops
 };

 % STATE EPC
 \node[state,text width=3cm,
  right of=PROP,
  node distance=4cm,
  anchor=center] (SEARCH) 
 {Search for more explicit DangerStates
 };

 \node[state,
  right of=SEARCH,
  node distance=4cm,
  text width=3.5cm,
  anchor=center] (CHECK2) 
 {\textbf{Check probability bounds:}
 is $P[\Diamond \neg\phi \mid I]>y$? 
 };

 \node [state, above of =CHECK2, node distance =2cm, xshift=0.7cm](YES2){PASS};
 \node [state, above of =CHECK2, node distance =2cm, xshift=-0.7cm](NO2){FAIL};

 \node [draw=blue, thick,dotted, fit= (YES)(NO) (CHECK), label={[xshift=1cm]early termination}] {};
  \node [draw=blue, thick,dotted, fit= (YES2)(NO2) (CHECK2), label={ final termination}] {};

 % draw the paths and and print some Text below/above the graph
 \path (REFINE) 	edge  (CHECK)
 (CHECK)     	edge [bend left] node [right]{Don't know}(LOOPS)
 (LOOPS)     	edge [bend left] node [left]{New loops}(CHECK)
 (LOOPS)     	edge node [right]{$\,\,$No new loops}(PROP)
 (CHECK)      edge node [sloped, above]{Yes}(YES)
 (CHECK)      edge node [sloped, below]{No}(NO)
 (PROP)   	edge node [sloped, above]{yes} (SEARCH)
 (PROP)     edge node [sloped, above] {No. $k++$} (REFINE)
 (SEARCH)       	edge [bend left] (CHECK2)
 (CHECK2)  	edge [bend left] (SEARCH)
  (CHECK2)      edge node [sloped, below]{Yes}(YES2)
 (CHECK2)      edge node [sloped, below]{No}(NO2)
 (INIT)  	edge  (REFINE);

\end{tikzpicture}
\end{center}
\caption{Overview of pIC3. \label{fig:block_diag}}
\end{figure}
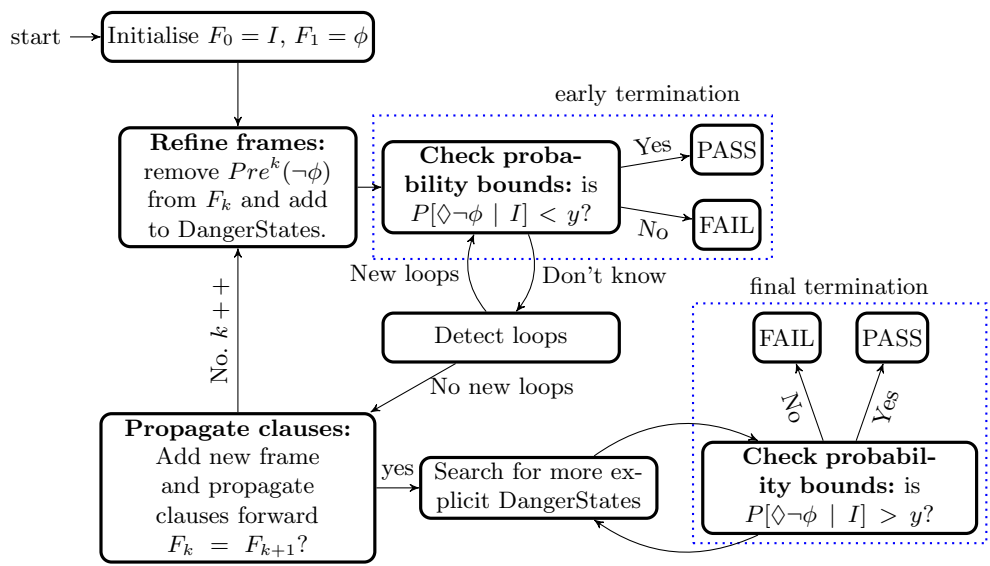
\end{center}

\section{Proof of correctness}
\label{sec:invariants}
%We will now show that pIC3 is sound and complete. 
%We begin by considering the nature of the $\DP$ for the property. 
%
\subsection{State space partitioning}
In Section~\ref{sec:alg} we have given an overview of how we partition the state
space using an IC3-like technique. 
We reach this state space partitioning by constructing regions $\SSt$ in the
same manner as the original IC3, 
and by defining $\DS$ to be $S \setminus (\neg \phi \cup \SSt)$. 
In order to build $\SSt$, we incrementally construct an invariant that 
%\red{[eventually]} 
implies $(\Post^*(I) \cap \neg \Pre^*(\neg \phi))$, 
by building a series of frames $F_0, ..., F_k$, where $\forall 0\leq i \leq k$ ,
 $(\Post^i(I) \setminus \Pre^i(\neg\phi)) \subset F_i $.  
To do this, the following program invariants must hold~\cite{DBLP:conf/vmcai/Bradley11}).
\begin{enumerate}
\item $I \implies F_0$
\item $F_i \implies F_{i+1} \,\,\,\,\, \forall \,1 \leq i \leq k$
\item $F_i \implies \phi \,\,\,\,\, \forall \,1 \leq i \leq k$
\item $F_i \wedge T \implies F'_{i+1} \,\,\,\,\, \forall \,1 \leq i \leq k$
\end{enumerate}

As we remove the states in $\Pre^n(\neg\phi)$, we add them to $\DS$. This gives
us a new program invariant, specific to pIC3. This ensures that $\DS$ only
contains states that are part of $\DP$ or that are unreachable from~$I$.
\begin{enumerate}
\setcounter{enumi}{4}
\item $\DS$$\subseteq \Pre^*(\neg\phi) \cap \phi$ and $\DS \cap F_i = \emptyset$
\end{enumerate}

The state space is correctly partitioned when we reach a frame closed under the
transition relation, i.e., $F_k = F_{k+1}$. Together, these four invariants and
reaching $F_k = F_{k+1}$ guarantee a correct state space partitioning. 

\subsection{Bounds on reachability probabilities}
In Section~\ref{sec:bounds}, we have given the formula for the upper and lower
bounds on the reachability probabilities for states in $\DS$, 
provided requirements for ``early termination'', 
and defined ``final termination'', which occurs after $F_k = F_{k+1}$. 
%The algorithm terminates and returns $\mathit{false}$ when an SMT solver can
%find no solutions to $L_p(I)<y$. Hence, it is necessary that $L_p$ is an under
%approximation, for this to be correct. The algorithm terminates and returns
%true when an SMT solver can find no solutions to $U_p(I)>y$, so it is necessary
%for $U_p$ to be an over approximation. We call both of these cases ``early
%termination''. The algorithm terminates and $L_p(I) = U_p(I) = Pr[\Diamond \neg
%\phi]$ when all states within $\DS$ have been found, which occurs when $F_k =
%F_{k+1}$, and all transitions within $\DS$ have been found. We call this
%``final termination''.
This gives us four more program invariants necessary for a correct result on
termination.
\begin{enumerate}
\setcounter{enumi}{5}
\item $L_p \leq Pr[\Diamond \neg \phi |s]: \forall s \in$ $\DS$
\item $U_p \geq Pr[\Diamond \neg \phi |s]: \forall s \in$ $\DS$
\item $L_p$ only ever increases
\item $U_p$ only ever decreases
\end{enumerate}

\subsubsection{Correctness}
\begin{proposition}
Altogether, program invariants (1) to (9) and termination yield total
correctness of the algorithm. 
Consequently, 
we show that our algorithm is correct by showing that the invariants hold
throughout the pseudo-code and that it terminates. 
The proof that the invariants hold can be found in Appendix A.
\end{proposition}

\subsubsection{Termination}

\begin{Lemma}
%Why is the loop guaranteed to terminate? 
Invariant (2), and the fact that $k$ is an index, dictate that the frame
sequence is non-decreasing. 
Then, due to the check $F_k = F_{k+1}$, the sequence must be strictly increasing
until termination. 
As the transition system is finite,  $F_k = F_{k+1}$ must be obtained in finite
time.%\red{in time that is linear with the state-space cardinality}.  
%after at most $\#states + 1$ steps.
\end{Lemma}
\begin{Lemma}
Once $F_k = F_{k+1}$, we build the full set of linear equations for the upper
and lower bounds in $\DS$. 
There are a finite number of states in $\DS$, and each state has a finite number
of transitions. 
Thus we can obtain a full set of linear equations in finite time. 
\end{Lemma}

\begin{proposition}
By Lemma 1 and Lemma 2 and the fact that a set of linear equations is solvable
in polynomial time, it is guaranteed that``final termination'' will occur in
finite time. 
\end{proposition}

Note that the algorithm will terminate ``early'' if the probability bounds show
that $Pr[\Diamond \neg \phi \mid I]$ is guaranteed to be on one side of the
threshold.

\section{Experimental results and discussion}
\label{sec:implement}
\subsection{Implementation}

We implement pIC3 in
C++, following the pseudocode in Appendix~\ref{sec:proof}. We use MiniSat 2.2.0 and 
Z3 4.4.2 for incremental SAT and SMT solving.

We input models directly as clausal transition relations, and do not explore the challenge of model encoding.
Our experiments have shown that the performance of pIC3 is highly dependent on the encoding of the model, with 
a poor encoding causing models to run as much as 10 times slower. We note that in order to run PRISM benchmarks it is possible to 
generate CNF from the BDD transition relations generated by PRISM, as in~\cite{DBLP:conf/cpaior/AbioGMS16, DBLP:conf/forte/BraitlingWBJA11}.

In order to assess the performance of pIC3 over varying model size, we construct
a model where we can increase the states without changing the regularity of the structure of the model (we know that the regularity
will affect both the performance of PRISM and pIC3). This model is
an aggregation of multiple models of Knuth's dice~\cite{KY76} with the properties that the probability of all, or
a set number, of the dice rolling a 
number 6 is below a certain threshold. 
We build identical models for PRISM and run the same experiments. 

We conducted the experimental evaluation on a 12-core 2.40 GHz Intel Xeon E5-2440 with 
96 GB of RAM and Linux OS. We limit experimental run time to 24 hours and use the Linux time command to measure this for pIC3.
We compare to PRISM 4.3.1, running the hybrid engine, and take the verification times for PRISM 
from the program output, omitting model construction time.%, and 
%we measure verification time for pIC3 using the linux time command. 
The properties we use are $P_{<y}[ F \neg\phi]$, 
and we run experiments on each model for a $y$ for which the property holds, and a $y$ for which the property does not hold.
We pick $y$ to be within $0.1$ of the actual probability of reaching $\neg\phi$.
We show the number of $\neg\phi$ states
in the table. 

%We run pIC3 on a selection of academic examples taken from the PRISM literature\cite{KY76,KNP12a}, and probabilistic
%adaptations of models taken from the IC3 literature~\cite{DBLP:conf/fmcad/SomenziB11}.  
%We run all models with a probabilistic bounded reachability property.
%We time both pIC3 and PRISM using the bash time command, and give the result as the user plus system time. We use the
%default hybrid engine in PRISM, which outperforms the other PRISM engines for our case studies.
\begin{table}[h]
\begin{center}
    \begin{tabular}{  | c | c | c | c | c | c | c|}
    \hline
    %5 dice, model 24
    states &  transitions      & $\neg\phi$ states  & \multicolumn{2}{|c|}{failing property}   & \multicolumn{2}{|c|}{passing property} \\
    	 &       	 &                     & pIC3	  & PRISM &  pIC3 & PRISM \\
     \hline
    371293   & 2353756 & 1         &    0.08s  &  0.22s & 0.92s & 0.21s  \\
     %        &           & 1       	   & 1e-5 (pass)      & 0.92s & 0.21s \\  
	         &           & 12           & 4.96s  & 0.21s &  6.43s &  0.21s \\
     	  %   &            & 12         &       pass           &  6.43s &  0.21s   \\ 
             & & 144  & 17.47s & 0.22s  & 1m39s & 0.22s  \\
           %  & & 144  & pass & 1m39s & 0.22s \\ 
     	     & & 1728  &  1m3s & 0.24s & 32m14s & 0.25s  \\ \hline
       %      &  & 1728  & pass  & 32m14s & 0.25s \\ \hline
    %6 dice

    4826809 & 35897772 & 1         &   0.7s  &  2.92s & 5.03s & 2.94ss\\
          %   &           & 1       	   & pass      & 5.03s & 2.94ss \\  
	         &           & 12          & 11.2s  & 2.82s   &  52.2s & 2.91s\\
     	   %  &            & 12         &       pass   &  52.2s & 2.91s   \\ 
             & & 144  & 1m46s & 2.83s  & 11m52s & 2.94s  \\
           %  & & 144  & pass & 11m52s & 2.94s \\ 
     	     & & 1728  & 165m6s  & 2.86s & 296m43s & 2.93s\\ \hline
            % &  & 1728  & pass  & 296m43s & 2.93s \\ \hline
    
    %7 dice

    62748517 & 24652041644 & 1       &   4.2s  &  39.4s & 44.7s & 39.0s \\
         %    &           & 1       	   & pass      & 44.7s & 16m1s \\  
	         &           & 12           & 25.5ss  & 40.2s  &  10m32s & 34.8s  \\
     	    % &            & 12         &       pass   &  10m32s & 16m1s   \\ 
             & & 144  & 11m58s & 40.0s & 155m35s & 40.5s   \\
          %   & & 144  & pass & 155m35s & 17m7s \\ 
     	     & & 1728 & TO  & 38.6s & TO & 38.5s \\ \hline
           %  &  & 1728  & pass  & TO & 18m53s \\ \hline
  
    %8 dice

    815730721 & 7837799824 & 1         &   30.9s  & 8m24s & 7m57s &  8m21s \\
       %      &           & 1       	   & pass      & 7m57s &  8m21s\\  
	         &           & 12           & 5m59s  & 8m20s &  145m31s & 8m14s  \\
     	   %  &            & 12         &       pass   &  145m31s & 8m14s   \\ 
             & & 144  & 203m31s &   8m15s & 1293m32s & 8m16s\\
           %  & & 144  & pass & 1293m32s & 8m16s \\ 
     	     & & 1728  & TO  & 8m9s & TO & 8m10s \\ \hline
            % &  & 1728  & pass  & TO & 8m10s \\ \hline
    
    %8 dice 815730721 states 7837799824 trans
    %9 dice 10604499373 states 113346216612trans

    10604499373 & 113346216612 & 1      &   5m47s  & X & 59m44s & X \\
         %    &           & 1       	   & pass      & 59m44s & X \\  
	         &           & 12       &  31m45s &  X  & 1279m10s  & X \\
     	   %  &            & 12         &       pass   & 1279m10s  & X  \\ 
             & & 144   & 1116m5s &  X & TO &  X  \\\hline
          %   & & 144  & pass & TO &  X \\ \hline

    \hline
    \end{tabular}
    \caption{Comparison of pIC3 and PRISM. X denotes an out-of-memory error, TO denotes a run-time exceeding 24 hrs. }
    \end{center}
\end{table}

Experimental results show that PRISM 
performs consistently well regardless of the number of bad states for models up to $10^9$ states,
and is equally fast to pass or fail a property, and generally faster than pIC3. 
However, pIC3 is faster than PRISM for large models with smaller numbers of $\neg\phi$ states, 
and can evaluate some models with $10^{10}$ states. 

This is because the performance of pIC3 is not dependent on the size of the model, but the number of CTIs that need 
to be removed. This is proportional to the number of $\DS$
required to prove or refute the property. 
However, as the number of $\neg\phi$ states and $\DP$ increases, the time taken
for pIC3 to prove or refute a property increases well beyond the time PRISM takes.
To improve scalability, we need to reduce the number of CTIs that must be removed. Generalising to find 
minimal inductive clauses from CTIs~\cite{DBLP:conf/fmcad/BradleyM07} will help this, but in future
work we plan to explore pre-processing steps towards counter-example minimisation, such as those used in~\cite{DBLP:conf/tacas/WimmerJABK12}.

In summary the experimental results show that, 
whilst pIC3 is not a fast choice for small systems where the set of $\DS$ is a large proportion of the total state space, 
it is able to tackle models where PRISM fails, and can be well-suited to the verification of rare-events.

\section{Related Work}
\label{sec:related_work}

Probabilistic IC3 is based on the IC3 algorithm proposed in~\cite{DBLP:conf/vmcai/Bradley11} 
%We recommend the tutorial by Bradley and
%Somenzi~\cite{DBLP:conf/fmcad/SomenziB11}, for an introduction to the
%algorithm. 
which tackles finite state non-probabilistic finite state systems. %, and the
%primary application area is hardware model checking
%
IC3 has been extended to SMT~\cite{DBLP:conf/sat/HoderB12}, 
and applied to infinite state systems~\cite{DBLP:conf/tacas/CimattiGMT14} using
predicate abstraction, implemented in NuXmv~\cite{DBLP:conf/cav/CavadaCDGMMMRT14}. 
The IC3 framework has, however, not yet been extended to probabilistic
models, which is the key contribution of this work. 

The current standard for probabilistic model checking is
PRISM~\cite{DBLP:conf/cav/KwiatkowskaNP11}. PRISM contains four main engines,
based on MTBDDs, BDDs, sparse matrices, and explicit-state methods. 
The default engine enabled is the hybrid engine, which uses a combination of
symbolic and explicit state data structures. The PRISM website states that PRISM 
can handle models 
of roughly $10^8$ states, depending on the structure of the model, and the number of 
distinct probabilities. 

 %The explicit engine does not scale well, but is good for models with a large
 %state space with only a small proportion of states being reachable. 
 %

SAT-based model checking techniques have been extended to probabilistic systems.
Stochastic SMT (SSMT) was first proposed by~\cite{DBLP:conf/hybrid/FranzleHT08},
who introduced randomised quantifiers to predicates in SMT. 
The original paper uses SSMT to consider bounded reachability in probabilistic
automata, which is an incomplete technique. 
The approach is, however, more general than that, and can be applied easily to
other properties, and has been integrated with Craig
Interpolation~\cite{DBLP:conf/rp/MahdiF14}. 
SSMT is a very different approach to pIC3, and builds on the idea of lazy form
of quantifier elimination to compute reachable sets. 
We have not compared performance of our approach to SSMT as the available SSMT
implementation uses the bounded model checking (BMC) framework and, as such, is
incomplete. 

Our work is closely related to work on generation of counter-examples for Markov chains.
The original work in this area~\cite{DBLP:conf/tacas/HanK07} applies k-shortest-path algorithms to Markov chains, and 
then computes a counterexample made up of the minimum number of paths. This method requires 
explicit state representation and so does not scale well to larger systems. 
Bounded Model Checking has been used to generate counter-examples for Markov chains~\cite{DBLP:conf/forte/BraitlingWBJA11}. 
As mentioned previously, there are several advantages to using IC3 over BMC: IC3 unrolls the transition relation only one step; it makes much more effective use of incremental SAT solving than BMC; and the reasoning in IC3 is localised and driven by the property being checked. 
The paper mentions obtaining incomplete assignments from a SAT solver, although it is not clear if
they implement this. This is similar to the generalisation procedure in pIC3. Their method requires computing
the length of the shortest counter-example path using the OBDDs, and so is limited to models that are
small enough for this to be possible. 

Similarly, \cite{DBLP:conf/tacas/WimmerJABK12} generate counter-examples for Markov chains
in the form of a minimal-critical subsystem. The subsystem can be considered similar to the $\DS$ in pIC3,
and they also use an SMT solver to solve a system of linear equations expressing
the reachability probability, but they solve the system for ALL states in the system and introduce boolean
variables for each state to identify the relevant states for the property.
pIC3 partitions out the relevant state-space using IC3-like techniques, and then solves a much 
smaller SMT problem.  
The authors also solve the same problem using Mixed Integer Linear Programming and introduce optimizations
that may benefit pIC3, such as introducing constraints based on SCCs. Counter example minimisation is tackled for Markov Chains \cite{DBLP:conf/hvc/AndresDR08} by abstracting away details of strongly connected components by replacing with edges that have a probability of walking through the SCC. It may be possible to apply this method
to pIC3 as a pre-processing step. 
%\cite{DBLP:conf/vmcai/DehnertKP13} use SMT for probabilistic bisimulations, effectively 
%shrinking the state-space that it is necessary to search.
% \cite{DBLP:conf/qest/DammanHK08} find counterexamples in DTMCs 
%using the shortest path algorithm, and then 
%represent these counterexamples as regular expressions. This is similar to ~\cite{DBLP:conf/ictac/Daws04}, 
%who use regular expressions to represent sets of paths and calculate the exact rational value of the probability
%measure in DTMC model checking. 

%Recent work has extended bounded model checking for probabilistic
%programs~\cite{DBLP:journals/corr/0001DKKW16}.
%
%

\section{Conclusions and further work}
\label{sec:extensions}

We have developed and implemented a probabilistic model checking algorithm
for bounded probabilistic reachability properties, based on incremental
construction of inductive invariants and a system of linear inequalities, 
and shown it is sound and complete. 
We have run pIC3 on a series of models of increasing size and found that pIC3 can verify rare events in systems too large
for PRISM to handle.

In future work we aim to increase the scalability to systems that require larger numbers of counter-examples
to prove or refute properties, by exploring techniques in~\cite{DBLP:conf/tacas/WimmerJABK12}, and by improving
the generalisation of CTIs.
Use of incremental SAT solving in IC3 is still an active area of research, and should be explored further for pIC3.
In addition, like the original
IC3, our algorithm is well-suited to parallel computation~\cite{DBLP:conf/vmcai/Bradley11}

\bibliography{pic3}{}
\bibliographystyle{plain}
\appendix
\section{Further proofs and pseudo-code}
\label{sec:proof}
We give pseudocode for our implementation of pIC3, and demonstrate
that the program invariants in Section~\ref{sec:invariants} hold in each part.

\subsubsection{main()}
\label{sec:main}

The top level function of pIC3 is shown in Figure~\ref{fig:block_diag}. In
addition to initialising the series of frames, pIC3 begins by initialising the
upper and lower probability bounds. We will look at the functions
extendFrontier, propagateClauses and checkTermination in more detail, but first
consider whether the program invariants hold after initialisation, at line
\ref{line:endinit}:
\begin{itemize}
\item invariant (1) holds due to $F_0 = I$
\item invariant (2) and (3) holds as, $I \implies \phi$ by the check in line 2,
and $F_1 = \phi$. 
\item invariant (4) is required to hold from frame $F_1$ onwards.
\item invariant (5) holds as $\DS$ is empty
\item invariants (6), (7) hold by the initialisation as 0 and 1
\item invariants(8), (9) hold as no changes have been made
\end{itemize}

\subsubsection{extendFrontier()}
\label{sec:extendfrontier}
Assume we are in iteration $k$ of the algorithm and frames $F_0, ..., F_k$
satisfy the program invariants (1)-(9). First a new frame $F_{k+1} = \phi$ is
added, as the largest possible over-approximation of states reachable in $k$
steps that does not violate $\phi$. Now we need to make the program invariants
hold again.
\begin{itemize}
\item invariant(1) is not affected
\item invariant(3) holds by construction
\item invariant (2) will hold in the end, since we will modify frames only by
adding the same clause to all frames $F_0, .., F_i$ for some $i$. 
\item Preserving invariant(4), $F_i \wedge T \implies F_{i+1}'$ becomes the
crucial part of this function: if invariant (4) holds, the function returns
without taking other actions. If it does not hold, it must be preserved by
removing the state in $F_k$ that has a transition that leads outside of
$F_{i+1}$. Thus the removeCTI function must remove
the counterexample to induction (CTI) whilst preserving the other program
invariants.
\item program invariants(5)--(9) must be preserved by UpdateProbability. 
\end{itemize}

%%%%%%%%%%%%%%%%%%%%%%%%%%%%%%%%
%%%%%%%%%%%%%%%%%%%%%%%%%%%%%%%%
\subsubsection{removeCTI()}
\label{sec:removeCTI}

The occurrence of a CTI $s$ in the frame $F_k$ can have 2 reasons:
\begin{itemize}
\item $s$ is reachable from $I$, and part of a counter-example path 
\item it is not yet discovered that $s$ is not reachable from $I$.  
\end{itemize}

We avoid differentiating between these cases, and remove $s$ from $F_k$ and add
$s$ to $\DS$ regardless. We pay a slight overhead cost for this of maintaining
an upper and lower bound for $\Prob [\Diamond \neg \phi \mid s]$ for unreachable
states that do not contribute to $\Prob [\Diamond \neg \phi \mid I]$. We remove
the states with backwards search. 

We build a queue of states to remove, and each entry $<q,i>$ tells us to remove
state $q$ from frame $F_i$. We first check that we cannot reach $q$ from $I$, if
we can then we have found a complete counter-example path and we add $q$ to
$\DS$ and return to the main function. If we cannot, then we add a clause $c
\implies \neg q$ to $F_1, ..., F_j$ where $F_j$ is the outermost frame where $c$
is inductive. Adding the clause removes $q$ from each frame. The clause $c$ is
an inductive generalisation of $\neg q$.

 If $j \geq i-1$, we have successfully removed $q$. If not, we get the
predecessor of $q$, add that to the queue of states to remove. We also add $q$
to the list of states to remove, as it may have other predecessors and not be
fully removed yet. We repeat this until the queue is empty. 

Now consider the program invariants at the end of removeCTI():

\begin{itemize}
\item invariant (1) holds as no changes are made to frame $F_0$
\item invariant (2) is not violated as clauses are added to all frames $F_0
...F_i$ for some $i$. 
\item invariant (3) holds as the formula for the frames are only ever extended,
and thus only ever become stricter.
\item invariant (4) is enforced as predecessors are revealed and enqueued to be
deleted and added to $\DS$. 
\item invariant(5) holds as states added to $\DS$ are found to be predecessors
to $\neg\phi$ and states are only added to $\DS$ as they are removed from the
frames.
\item for program invariants (6) to (9) we must look at the updateProbability
function.
\end{itemize}

%%%%%%%%%%%%%%%%
\subsubsection{updateProbability()}
\label{sec:updateProbability}
The updateProbability function incrementally constructs the upper and lower
bounds for $\Prob[\Diamond \neg \phi \mid s]$ for all $s \in \DS$.
\begin{itemize}
\item program invariants (1) to (5) are unaffected
\item program invariants (6) and (8) are true. We add transitions as we find
them to the upper and lower bounds, building a subset of transitions that all
lead to $\neg\phi$. All states in $\DS$ are connected to $\neg\phi$ by this
chain of transitions, and so there are no bottom strongly connected components,
and the upper and lower bounds have a unique solution. We know the sum of
the transition probabilities of the transitions we have not found yet, and for
the upper bound we assume all these transitions connect to states in $\neg\phi$,
so have a probability mass of 1. For the lower bound we assume all these
transitions connect to states with a probability mass of 0. The check in line
\ref{line:check} ensures we only add each transition once. We only add to $L_p$,
and so $L_p$ can only increase.
\item program invariants (7) and (9) are true for the above reasons. We only
subtract from $U_p$.
\end{itemize}

\subsubsection{propagateClauses()}
\label{sec:propagateClauses}

When we add a new frame, we propagate forward all clauses that are inductive
from the previous frames. The propagateClauses() function preserves invariant
(4) by line \ref{line:inv4}. It preserves invariant (2), as clauses are added
that have already been added to predecessor frames. The other program invariants
are unaffected.

%%%%%%%%%%%%%%%%%%%%
%%%%%%%%%%%%%%%%%%%%
%%%%%%%%%%%%%%%%%%%%
%%%%%%%%%%%%%%%%%%%%
%%%%%%%%%%%%%%%%%%%%%%%%%%%%%%%%%%
%%%%%%%%%%%%%%%%%%%%%%%%%%%%%%%%%%
\subsubsection{checkTermination()}
\label{sec:termination}
The checkTermination() and findLoops() functions ensure that, if we reach $F_k =
F_{k+1}$ without terminating, we have all transitions in $\DS$ included in $L_p$
and $U_p$. Once all transitions are found, we have a full set of linear
equations to be solved and a precise answer for $L_p(I) = P[\Diamond \neg
\phi]$.  

\begin{algorithm}[]
\begin{algorithmic}[1]
\Function{pIC3}{}
	%\State add $I$, %Pre-initial state with transition to $I$ 
	\If {UNSAT($I \implies \phi)$}
		\State property is FALSE
	\EndIf	
	\State $F_0 := I, F_1 :=\phi, k:=1$
	\State $\forall s \in \neg \phi: L_p = 1 $
	\State $ \forall s \in \phi: L_p = 0$
	\State $ \forall s \in \phi: U_p = 1$ \label{line:endinit}
	\While{true}
		\State extendFrontier() \label{line:extfront}
		\State propagateClauses() \label{line:propclause}
		\State checkTermination() \label{line:checkterm}
		\State $k:=k+1$
  	\EndWhile
\EndFunction
\end{algorithmic}
\label{alg:pic3_main}
\end{algorithm}

\begin{algorithm}[]
\begin{algorithmic}[1]
\Function{extendFrontier}{}
	\State $F_{k+1} := \phi$
	\While{$SAT(F_k \wedge T \wedge \neg \phi')$}		
	\State s:= predecessor of $\neg \phi$ extracted from SAT witness
	\State s':= successor of s extracted from SAT witness
	\State add (s,s') to $\DS$
	\State updateProbability(s,s')
	\State removeCTI(s)			
  	\EndWhile
\EndFunction
\end{algorithmic}
\label{alg:extendFrontier}
\end{algorithm}

%%%%%%%%%%%%%%%%%%%%%%%%%%%%%%%%
%%%%%%%%%%%%%%%%%%%%%%%%%%%%%%%%
%%%%%%%%%%%%%%%%%%%%%%%%%%%%%%%%
\begin{algorithm}[]
\begin{algorithmic}[1]
\Function{removeCTI}{}
	\State $states = \{\langle s,k \rangle \}$ 
	\While{$states$ not empty}
		\State$\{ \langle q,i \rangle \}$ = pop element of $states$ 
\Comment{the inner-most}
		\If {$SAT(F_0 \wedge T \wedge \neg q \wedge q')$}
			\State updateProbabilty($I$,$q$) 
			\State $F_i := F_i \wedge \neg q$  \Comment{remove $q$, not $I$}
			\State \textbf{break}
		\EndIf	

		\State $j$:= maximal $j$ for which $SAT(F_j \wedge T \wedge \neg q
\wedge q'$)
		\State c:= inductiveGeneralisation($\neg q$)
		\For {$l$ from 0 to $j+1$}
			\State $F_l := F_l \wedge c$
		\EndFor
		\If{$j \geq i-1$}
			\State \textbf{break} \Comment{have removed state with no reachable
predecessors}
		\EndIf
		\State w:= witness for $SAT(F_{j+1} \wedge T \wedge \neg q \wedge q')$
		\State t:= predecessor of q, extracted from w
		\State add (t,q) to $\DS$
		\State updateProbability(t,q);
		\State $states := states\: \cup \langle t, j+1 \rangle$
		\State $states := states\: \cup \langle q, i \rangle$	
			
  	\EndWhile

\EndFunction
\end{algorithmic}
\label{alg:removeCTI}
\end{algorithm}
%\todo{take some detail from removeCTI}
%
%
%%%%%%%%%%%%%%%%%
%%%%%%%%%%%%%%%%%
%%%%%%%%%%%%%%%%%
%%%%%%%%%%%%%%%%%
\begin{algorithm}[]
\begin{algorithmic}[1]
\Function{updateProbability}{$s$, $s'$}
	\If{$T(s,s') \not\in L_p(s $} \Comment{if we haven't already found this
transition} \label{line:check}
	\State $L_p(s) = L_p(s) + T(s,s')*L_p(s')$ \Comment{Append transition to
$L_p$}
	\State $U_p(s) = U_p(s) - T(s,s')$
	\EndIf
	%\If{$s==I$}
	%	\State remove transition $T(I*,s)$ from $T$ and $T2$\Comment{stop this
	%path being found again}		
	%\EndIf	
	\If {$\neg SAT(L_p(I)<y)$} \Comment{SMT solver}
			\State \Return FAIL
	\EndIf 
	\If {$\neg SAT(U_p(I)>y)$} \Comment{SMT solver}
			\State \Return PASS
	\EndIf 
\EndFunction
\end{algorithmic}
\label{alg:updateProbability}
\end{algorithm}

\begin{algorithm}[]
\begin{algorithmic}[1]
\Function{propagateClauses}{}
	\For {i = 1 to k}
		\For {every clause $c \in F_i$}
			\If{not $SAT(F_i \wedge T \wedge \neg c')$} \label{line:inv4}
				\State $F_{i+1} = F_{i+1} \wedge c$
			\EndIf
		\EndFor
	\EndFor			
\EndFunction
\end{algorithmic}
\label{alg:propagateClauses}
\end{algorithm}

%%%%%%%%%%%%%%
%%%%%%%%%%%%%%
%%%%%%%%%%%%%%
\begin{algorithm}[]
\begin{algorithmic}[1]
\Function{checkTermination}{}
	\If{$F_i = F_{i+1}$ for some i}
		\State findLoops()
		\State \Return TRUE
	\EndIf		
\EndFunction
\end{algorithmic}
\label{alg:checkTermination}
\end{algorithm}

%%%%%%%%%%%
\begin{algorithm}[]
\begin{algorithmic}[1]
\Function{findLoops()}{}
\State $\mathit{foundTransitions} = (\DS, \DS')$ \Comment{every transition
already found in $\DS$}
	\While{$SAT((\DS) \wedge T \wedge \neg \mathit{foundTransitions} \wedge
\DS')$}
		\State s:=predecessor of $\DS'$ extracted from SAT witness 
		\State updateProbability(s,s')
		\State add $T(s,s')$ to $\mathit{foundTransitions}$ 
	\EndWhile		
\EndFunction
\end{algorithmic}
\label{alg:findLoops}
\end{algorithm}

\end{document}